\newtheorem{theorem}{Theorem}

\newtheorem{definition}[theorem]{Definition}

\newtheorem{proposition}[theorem]{Proposition}
\newtheorem{remark}[theorem]{Remark}

\documentclass[12pt]{article}%
\usepackage{amssymb}
\usepackage{amsmath}%
\usepackage{amsfonts}%
\usepackage{graphicx}

\begin{document}

\title{INFORMATION DYNAMICS AND\ ITS APPLICATION TO RECOGNITION PROCESS}
\author{Masanori Ohya\\Department of Information Sciences,\\Science University of Tokyo\\278 Noda City, Chiba, Japan}
\date{}
\maketitle

\section{Introduction}

Professor H. Ezawa has worked on vast area related with fundamental physics.
His interests and pioneering works are not limited to
''elementary-particle-oriented'' works but also cover more \textit{complex}
system, such as finite temparature quantum field theory and nonequilibrium
theory. It is, therefore, great honor of mine to contribute to this volume by
presenting an attempt, information dynamics, to treat various complex systems
mathematically and its one of the latest application, a description of
recognition process\cite{FFO} based on the works \cite{FO1,FO2}. In this
section we give a review on what is the complex system. The discussion leads
the introduction of Information Dynamics in natural way.

The complex system has been considered in Santafe research center as follows:

(1) A system is composed of several elements called agents. The size of the
system (the number of the elements) is medium.

(2) The agent has intellegence.

(3) Each agent has interaction due to local information. The decision of each
agent is determined by not all information but the limited information of the system.

Under a small modification, I define the complex system as follows:

(1) A system is composed of several elements. The scale of the system is often
large but not always, in some cases one.

(2) Some elements of the system have special (self) interactions (relations),
which produce a dynamics of the system.

(3) The system shows a particular character (not sum of the characters of all
elements) due to (2).

\begin{definition}
A system having the above three properties is called ``complex system''. The
''complexity''of such a complex system is a quantity measuring that
complexity, and its change describes the appearance of the particular
character of the system.
\end{definition}

There exist such measures describing the complexity for a system, for
instance, variance, correlation, level - statistics, fluctuation, randomness,
multiplicity, entropy, fuzzy, fractal dimension, ergodicity (mixing, flow),
bifurcation, localization, computational complexity (Kolmogorov's or
Chaitin's), catastrophy, dynamical entropy, Lyapunov exponent, etc. These
quantities are used case by case and they are often difficult to compute.
Moreover, the relations among these are lacking (not clear enough). Therefore
it is important to find common property or expression of these quantities. In
this paper, we introduce such a common degree to describe the chaotic aspect
of quantum dynamical systems. Further we describe the function of barin in the
framework of information dynamics \cite{O3}(ID for short) and we discuss the
value of information attached to the brain in terms of the complexity in ID
and the chaos degree\cite{O7,O9}.

\section{Information Dynamics}

There are two aspects for the complexity, that is, the complexity of a state
describing the system itself and that of a dynamics causing the change of the
system (state). The former complexity is simply called the ''complexity'' of
the state, and the later is called the ''chaos degree'' of the dynamics in
this paper. Therefore the examples of the complexity are entropy, fractal
dominion, and those of the chaos degree are Lyapunov exponent, dynamical
entropy, computational complexity. Let us discuss a common quantity measuring
the complexity of a system so that we can easily handle. The complexity of a
general quantum state was introduced in the frame of ID \cite{O3,IKO} and the
quantum chaos degree was defined in \cite{IKO2}, which we will review in this section.

Information Dynamics is a synthesis of dynamics of state change and complexity
of state. More precisely, let $(\mathcal{A},\mathfrak{S},\alpha(G))$ be an
input (or initial) system and $(\overline{\mathcal{A}},\overline
{{\mathfrak{S}}},\overline{\alpha}(\overline{G}))$ be an output (or final)
system. Here $\mathcal{A}$ is the set of all objects to be observed and
$\mathfrak{S}$ is the set of all means for measurement of $\mathcal{A}$,
$\alpha(G)$ is a certain evolution of system. Once an input and an output
systems are set, the situation of the input system is described by a state, an
element of $\mathfrak{S}$ $,$ and the change of the state is expressed by a
mapping from $\mathfrak{S}$ to $\overline{{\mathfrak{S}}}$, called a channel,
$\Lambda^{*}:\mathfrak{S}\to\overline{{\mathfrak{S}}}$ . Often we have
$\mathcal{A}=\overline{\mathcal{A}}$, $\mathfrak{S}=\overline{{\mathfrak{S}}}%
$, $\alpha=\overline{\alpha},$ which is assumed in the sequel. Thus we claim

\begin{center}
[Giving a mathematical structure to input and output triples

$\equiv$ Having a theory]
\end{center}

\smallskip For instance, when $\mathcal{A}$ is the set $M(\Omega)$ of all
measurable functions on a measurable space $(\Omega,\mathcal{F})$ and
$\mathfrak{S}(\mathcal{A})$ is the set $P(\Omega)$ of all probability measures
on $\Omega$ , we have usual probability theory, by which the classical
dynamical system is described. When $\mathcal{A}$ = $B(\mathcal{H}),$ the set
of all bounded linear operators on a Hilbert space $\mathcal{H}$, and
$\mathfrak{S}(\mathcal{A})$ = $\mathfrak{S}(\mathcal{H})$ , the set of density
operators on $\mathcal{H}$, we have a usual quantum dynamical system. In this
paper, we assume that both the input and output triple $(\mathcal{A}%
,\mathfrak{S},\alpha(G))$ is a C*-dynamical system or the usual quantum system
as above, and a channel, $\Lambda^{*}:\mathfrak{S}\to\mathfrak{S}$ is a
completely positive map.

There exist two complexities in ID, which are axiomatically given as follows:

Let $(\mathcal{A}_{t},\mathfrak{S}_{t},\alpha^{t}(G^{t}))$ be the total system
of both input and output systems; $\mathcal{A}_{t}\equiv\mathcal{A\otimes A}
,\mathfrak{S}_{t}\equiv\mathfrak{S\otimes S},\alpha^{t}\equiv\alpha
\otimes\alpha$ with suitable tensor products $\otimes.$ Further, let $C\left(
\varphi\right)  $ be the complexity of a state $\varphi\in$ $\mathfrak{S}$ and
$T\left(  \varphi;\Lambda^{*}\right)  $ be the transmitted complexity
associated with the state change$\;\varphi\to\Lambda^{*}\varphi.$ These
complexities $C$ and $T$ are the quantities satisfying the following conditions:

\begin{enumerate}
\item[(i)] For any $\varphi\in\mathfrak{S}$,
\[
C(\varphi)\ge0,\ T(\varphi;\Lambda^{*})\ge0.
\]

\item[(ii)] For any orthogonal bijection $j:ex\mathfrak{S}\mathcal{\rightarrow
}ex\mathfrak{S}$ ( the set of all extreme points in $\mathfrak{S} $ ),
\[
C(j(\varphi))=C(\varphi),
\]
\[
T(j(\varphi);\Lambda^{*})=T(\varphi;\Lambda^{*}).
\]

\item[(iii)] For $\Phi\equiv\varphi\otimes\psi\in\mathfrak{S}_{t}$,
\[
C(\Phi)=C(\varphi)+C(\psi).
\]

\item[(iv)] For any state $\varphi$ and a channel $\Lambda^{*},$%
\[
T(\varphi;\Lambda^{*})\le C(\varphi).
\]

\item[(v)] For the identity map ``id'' from $\mathfrak{S}$ to ${\ }%
\mathfrak{S}$.
\[
T(\varphi;id)=C(\varphi).
\]

\end{enumerate}

\begin{definition}
\noindent:Quantum Information Dynamics (QID) is defined by
\begin{align}
\left(  \mathcal{A},\mathfrak{S},\alpha(G);\;\Lambda^{*};\;C(\varphi
),T(\varphi;\;\Lambda^{*})\right)
\end{align}
and some relations R among them.

\end{definition}

There are several examples of the above complexities $C$ and $T$ such as
quantum entropy and quantum mutual entropy \cite{O1,O6}. Information Dynamics
can be applied to the study of chaos in the following sense:

\begin{definition}
\cite{O7,O9,IKO}$\psi$ is more chaotic than $\varphi$ as seen from the
reference system $\mathcal{S}$ if $C(\psi)\ge C(\varphi)$.

When $\varphi$ changes to $\Lambda^{*}\varphi$, the\textit{\ degree of chaos}
associated to this state change(dynamics) $\Lambda^{*}$ is given by
\[
D(\varphi;\Lambda^{*})=\inf\left\{  \int_{\mathfrak{S}}C(\Lambda^{*}%
\omega)d\mu;\mu\in M\left(  \varphi\right)  \right\}  ,
\]
where $\varphi=\int_{\mathfrak{S}}\omega d\mu$ is a maximal extremal
decomposition of $\varphi$ and $M\left(  \varphi\right)  $ is the set of such
measures. In some cases such that $\Lambda^{*}$ is linear, this chaos degree
$D(\varphi;\Lambda^{*})$ can be written as $C(\Lambda^{*}\varphi
)-T(\varphi;\;\Lambda^{*}).$
\end{definition}

Since ID has hierarchy (hierarchical structure), it can be applied several
open systems. Later we apply ID to Brain Dynamics.

\section{Entropic Chaos Degree (ECD)}

In the context of information dynamics, a chaos degree associated with a
dynamics in classical systems was introduced in \cite{O7}. It has been applied
to several dynamical maps such logistic map, Baker's transformation and
Tinkerbel map with succesful explainations of their chaotic characters
\cite{IOS}. This chaos degree has several merits compared with usual measures
such as Lyapunov exponent.

Here we discuss the quantum version of the classical chaos degree, which is
defined by quantum entropies in Section 2, and we call the quantum chaos
degree the entropic quantum chaos degree. In order to contain both classical
and quantum cases, we define the entropic chaos degree (ECD) in C*-algebraic
terninology. This setting will not be used in the sequel application, but for
mathematical completeness we first discuss the C*-algebraic setting.

Let $(\mathcal{A},\mathfrak{S})$ be an input C* system and $(\overline
{\mathcal{A}},\overline{{\mathfrak{S}}})$ be an output C* system; namely,
$\mathcal{A}$ is a C* algebra with unit $I$ and $\mathfrak{S}$ is the set of
all states on $\mathcal{A}$. We assume $\overline{\mathcal{A}}=\mathcal{A}$
for simlicity. For a weak* compact convex subset $\mathcal{S}$ (called the
reference space) of $\mathfrak{S}$, take a state $\varphi$ from the set
$\mathcal{S}$ and let%

\[
\varphi=\int_{\mathcal{S}}\omega d\mu_{\varphi}
\]
be an extremal orthogonal decomposition of $\varphi$ in$\mathcal{\ S}$, which
describes the degree of mixture of $\varphi$ in the reference space
$\mathcal{S}$ \cite{O2,OP}. The measure $\mu_{\varphi}$ is not uniquely
determined unless $\mathcal{S}$ is the Schoque simplex, so that the set of all
such measures is denoted by $M_{\varphi}\left(  \mathcal{S}\right)  .$ The
entropic chaos degree with respect to $\varphi\in\mathcal{S}$ and a channel
$\Lambda^{*}$ is defined by%

\[
D^{\mathcal{S}}\left(  \varphi;\Lambda^{*}\right)  \equiv\inf\left\{
\int_{\mathcal{S}}S^{\mathcal{S}}\left(  \Lambda^{*}\varphi\right)
d\mu_{\varphi};\mu_{\varphi}\in M_{\varphi}\left(  \mathcal{S}\right)
\right\}  \mbox{ }\left(  3.1\right)
\]
where $S^{\mathcal{S}}\left(  \Lambda^{*}\varphi\right)  $ is the mixing
entropy of a state $\varphi$ in the reference space $\mathcal{S}$
\cite{O4,IKO}. When $\mathcal{S=}\mathfrak{S,}$ $D^{\mathcal{S}}\left(
\varphi;\Lambda^{*}\right)  $ is simply written as$D\left(  \varphi
;\Lambda^{*}\right)  .$ This $D^{\mathcal{S}}\left(  \varphi;\Lambda
^{*}\right)  $ contains both the classical chaos degree and the quantum one.

In usual quantum system including classical discrete system, $\mathcal{A}$ is
the set $\mathbf{B}\left(  \mathcal{H}\right)  $ of all bounded operators on a
Hilbert space $\mathcal{H}$ and $\mathfrak{S}$ is the set $\mathfrak{S}%
\mathcal{(H)}$ of all density operators on $\mathcal{H}$, in which an extreme
decomposition of $\rho\in\mathfrak{S}\mathcal{(H)}$ is a Schatten
decomposition $\rho=\sum_{k}p_{k}E_{k}$ (i.e., $\left\{  E_{k}\right\}  $ are
one dimensional orthogonal projections with $\sum E_{k}=I),$ so that the
entropic chaos degree is written as%

\begin{align}
D\left(  \rho;\Lambda^{*}\right)  \equiv\inf\left\{  \sum_{k}p_{k}%
S(\Lambda^{*}E_{k});\left\{  E_{k}\right\}  \right\}  ,
\end{align}
where the infimum is taken over all possible Schatten decompositions and $S$
is von Neumann entropy. Note that in classical discrete case, the Schatten
decomposition is unique $\rho=\sum_{k}p_{k}\delta_{k}$ with the delta measure
$\delta_{k}\left(  j\right)  \equiv\left\{
\begin{array}
[c]{ll}%
1 & \left(  k=j\right) \\
0 & \left(  k\neq j\right)
\end{array}
\right.  ,$ and the entropic chaos degree is written by%

\begin{align}
D\left(  \varphi;\Lambda^{\ast}\right)  =\sum_{k}p_{k}S(\Lambda^{\ast}%
\delta_{k}),
\end{align}
where $\rho$ is the probability distribution of the orbit obtained from a
dynamics of a system and the channel $\Lambda^{\ast}$ is generated from the dynamics.

We can judge whether the dynamics $\digamma^{*}$ causes a chaos or not by the
value of D as%

\begin{align*}
D  & >0\mbox{ and not constant}\Longleftrightarrow\mbox{chaotic,}\\
D  & =\mbox{constant}\Longleftrightarrow\mbox{weak stable,}\\
D  & =\mbox{0}\Longleftrightarrow\mbox{stable.}
\end{align*}

\noindent The classical version of this degree was applied to study the
chaotic behaviors of several nonlinear dynamics \cite{IOS,O7}. The quantum
entropic chaos degree is applied to the analysis of quantum spin
system\cite{IKO2} and quantum Baker's type transformation\cite{IOV}, and we
could measure the chaos of these systems. The information theoretical meaning
of this degree was explained in \cite{O9}.

\noindent\qquad The ECD can resolve some inconvenient properties of the
Lyapunov exponent, another degree of chaos \cite{IOS,KOT}:

\begin{itemize}
\item[(1)] Lyapunov exponent takes negative value and sometimes $-\infty$, but
the ECD is always positive for any $a\geq0$.

\item[(2)] It is difficult to compute the Lyapunov exponent for some maps like
Tinkerbell map $f$ because it is difficult to compute $f^{n}$ for large $n$.
On the other hand, the ECD of $f\ $is easily computed.

\item[(3)] Generally, the algorithm for the ECD is much easier than that for
the Lyapunov exponent.
\end{itemize}

\section{Quantum Information Dynamic Description of Brain}

The Information Dynamics can be employed to describe not only several
classical and quantum physical physics but also life sciences. We will
construct a model describimg the function of brain in the context of Quantum
Inforamtion Dynamics (QID).

We study a possible function of brain, in particular, we try to describe
several aspects of the process of recognition. In order to understand the
fundamental parts of the recognition process, the quantum teleportation
scheme\cite{FO2} seems to be useful. We consider a channel
expression of the teleportation process that serves for a simplified
description of the recognition process in brain.

It is the processing speed that we take as a particular character of the
brain, so that the high speed of processing in the brain is here supposed to
come from the coherent effects of substances in the brain like quantum
computer, as was pointed out by Penrose. Having this in our mind, we propose a
model of brain describing its function as follows:

The brain system $\emph{BS}$ =$\mathfrak{X}$ is supposed to be described by a
triple ( $B(\mathcal{H)}$, $\mathcal{S}(\mathcal{H)}$, $\Lambda^{\ast}(G) $ )
on a certain Hilbert space $\mathcal{H}$ where $B(\mathcal{H)}$is the set of
all bounded operators on $\mathcal{H}$, $\mathcal{S}(\mathcal{H)}$ is the set
of all density operators and $\Lambda^{\ast}(G)$ is a channel giving a state
change with a group $G$.

Further we assume the following:

(1) $\emph{BS}$ is described by a quantum state and the brain itself is
divided into several parts, each of which corresponds to a Hilbert space so
that $\mathcal{H}$ =$\oplus_{k}\mathcal{H}_{k}$ and $\varphi=\oplus
_{k}\mathcal{\varphi}_{k},$ $\mathcal{\varphi}_{k}\in\mathfrak{S}%
(\mathcal{H}_{k}\mathcal{)}$. However, in this paper we simply assume that the
brain is in one Hilbert space $\mathcal{H}$ because we only consider the basic
mechanism of recognition.

(2) The function (action) of the brain is described by a channel
$\Lambda^{\ast}$=$\oplus_{k}\Lambda_{k}^{\ast}$. Here as in (1) we take only
one channel $\Lambda^{\ast}.$

(3) $\emph{BS}$ is composed of two parts; information processing part ''$P$''
and others ''$O$'' (consciousness, memory, recognition) so that $\mathfrak{X=}%
\mathfrak{X}_{P}\otimes\mathfrak{X}_{O}$, $\mathcal{H}$ =$\mathcal{H}%
_{P}\mathcal{\otimes H}_{O}$.

Thus in our model the whole brain may be considered as a parallel quantum
computer, but we here explain the function of the brain as a quantum
computer, more precisely, a quantum communication process with entanglements
like in a quantum teleportation process. We will explain the mathematical
structure of our model.

Let $s=\left\{  s^{_{1}},s^{_{2}},\cdots,s^{n}\right\}  $ be a given (input)
signal (perception) and $\overline{s}=\left\{  \overline{s}^{_{1}}%
,\overline{s}^{_{2}},\cdots,\overline{s}^{n}\right\}  $ the output signal.
After the signal $s$ enters the brain, each element $s^{j}$ of $s$ is coded
into a proper quantum state $\rho^{_{j}}\in\mathcal{S}\left(  \mathcal{H}%
_{P}\right)  ,$ so that the state corresponding to the signal $s$ is
$\rho=\otimes_{j}\rho^{_{j}}.$ This state may be regarded as a state processed
by the brain and it is coupled to a state $\rho_{O}$ stored as a memory
(pre-conciousness) in brain. The processing in the brain is expressed by a
properly chosen quantum channel $\Lambda^{\ast}$ (or $\Lambda_{P}^{\ast
}\otimes$ $\Lambda_{O}^{\ast})$. The channel is determined by the form of the
network of neurons and some other biochemical actions, and its function is
like a (quantum) gate in quantum computer\cite{O1,OV1}. The outcome state
$\overline{\rho}$ contacts with an operator $F$ describing the work as noema
of consciousness (Husserl's noema), after the contact a certain reduction of
state is occured, which may correspond to the noesis (Husserl's) of
consciousness. A part of the reduced state is stored in brain as a memory. The
scheme of our model is represented in the following figure.%

\begin{center}
\includegraphics[
height=4.8611in,
width=5.0142in
]%
{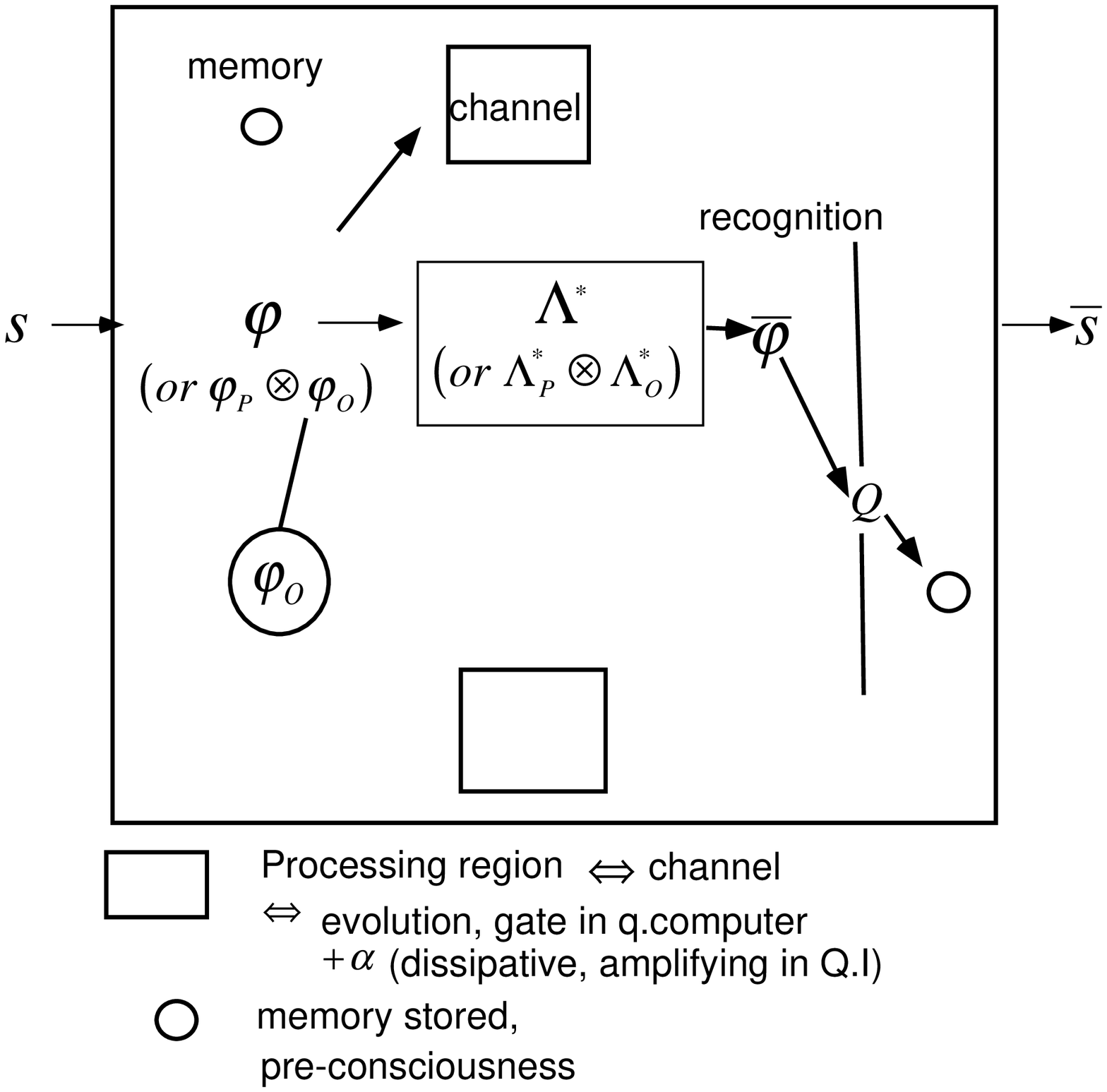}%
\end{center}

\begin{center}
{\large Figure of BRAIN}
\end{center}

\section{Value of Information in Brain}

The complex system responses to the information and has a particular role to
choose the information (value of information). Brain selects some information
(inputs) from huge flow of information (inputs). It will be important to find
a rule or rules of such selection mechanisum. In the model of Sec.4, an output
signal $s$ (information) is somehow coded into a quantum state $\varphi$, then
it runs in brain with a certain processing effect $\Lambda^{\ast}$ and a
memory stored, and it changes its own figure. Thus we have two standpoints to
catch the value of information in brain. Suppose that we have a fixed purpose
(intention) described by an operator $Q,$ then one view of the value of
information is whether the signal $s$ is important for the purpose $Q$ and the
processing $\Lambda^{\ast}$ and another is whether the processing
$\Lambda^{\ast}$ chosen in brain is effective for $s $ and $Q.$ From these
considerations, that value should be estimated by a function of the state
$\varphi\otimes\varphi_{O},$ a channel $\Lambda^{\ast}$ and an operator $Q,$
so that one possibility to define a measure $V(\varphi\otimes\varphi
_{O},\Lambda^{\ast},Q)$ estimating the effect of a signal and a function of
brain is as follows:

Define%

\[
V\left(  \varphi\otimes\varphi_{O},\Lambda^{*},Q\right)  =tr\Lambda^{*}%
\varphi\otimes\varphi_{O}Q
\]

\begin{definition}
Value of Information:
\end{definition}

\begin{itemize}
\item[(1)] $s=\left\{  s^{_{1}},s^{_{2}},\cdots,s^{n}\right\}  $ is more
valuable than $s^{^{\prime}}=\left\{  s^{\prime_{1}},s^{^{\prime}2}%
,\cdots,s^{^{\prime}n}\right\}  $for $\Lambda^{\ast}\ $and $Q$ iff
\[
V(\varphi\otimes\varphi_{O},\Lambda^{\ast},Q)\geqq V(\varphi^{^{\prime}%
}\otimes\varphi_{O},\Lambda^{\ast},Q).
\]

\item[(2)] $\Lambda^{\ast}$ is more valuable than $\Lambda^{^{\prime}\ast}$
for given $s=\left\{  s^{_{1}},s^{_{2}},\cdots,s^{n}\right\}  $ and $Q$ iff
\[
V(\varphi\otimes\varphi_{O},\Lambda^{\ast},Q)\geqq V(\varphi\otimes\varphi
_{O},\Lambda^{^{\prime}\ast},Q).
\]

\end{itemize}

The details of this estimator is discussed in \cite{FFO2}, where there exist
some relations between the information of value and the complexity or the
chaos degree under properly chosen complexity $C$ and transmitted complexity
$T.$ For instance, with entropy type complexities and a certain $Q$, we
conjecture (partially proved so far)%

\[
D\left(  \varphi\otimes\varphi_{O},\Lambda^{*};Q\right)  \leq D\left(
\varphi\otimes\varphi_{O},\Lambda^{^{\prime}*};Q\right)  \Longleftrightarrow
V\left(  \varphi\otimes\varphi_{O},\Lambda^{*},Q\right)  \geq V\left(
\varphi\otimes\varphi_{O},\Lambda^{^{\prime}*},Q\right)
\]
This result is quite natural because the more chaos a processing produces, the
less value it has$.$

\section{A Speculation of Brain Function}

The set of neurons in brain is divided into several parts and each part
corresponds to a configuration domain $G$, each point in which has two states,
excited or not. Thus $G\equiv\cup_{k}G_{k}$ with $G_{k}\cap G_{j}=\emptyset$
for any $k\neq j.$ Let assume that the Hilbert space $\mathcal{H}$ describing
the barin is Fock space on the square integrable random variables $L_{2}%
(G,\mu)$ with the counting measure $\mu,$ so that the whole Hilbert space
$\mathcal{H}$ is decomposed as%

\[
\mathcal{H\equiv\Gamma}\left(  L_{2}(G,\mu)\right)  =\otimes_{k}%
\mathcal{\Gamma}\left(  L_{2}(G_{k},\mu)\right)  .
\]
Let $\left\{  x_{1},x_{2},\cdots,x_{n}\right\}  $ describes the (positions of)
excited neurons in as certain domain $G_{k},$ so that the vector in
$L^{2}(G_{k},\mu)$ corresponding this configuration is denoted by $\sum
_{j=1}^{n}\delta_{x_{j}}$ by the delta measure $\delta_{x}$ corresponding to
$x.$

When we consider only one domain, for simplicity, denoted by the same $G$ and
it is decomposed ito the processing part $G^{P}$ and other part $G^{O}$
including the effect of conciousness as in the previous section, our Hilbert
space of the brain is $\mathcal{H\equiv\Gamma}\left(  L_{2}(G,\mu)\right)
=\Gamma\left(  L_{2}(G^{P},\mu)\right)  \otimes\Gamma\left(  L_{2}(G^{O}%
,\mu)\right)  .$ Along the above settings we may explain some functions of
brain in the terminologies of Fock space and quantum
teleportation\cite{FO1,FO2}, on which we are working now \cite{FFO2}.

In the sequel, we will explain the first trial explaining the brain function,
in particular the memory change due to recognition, based on the quantum
teleportation scheme done in \cite{FFO}.

Let us assume the Hilbert space $\mathcal{H}_{O}$ is composed of two parts,
before and after recognition. For notational simplicity, we denote the Hilbert
spaces by $\mathcal{H}_{1},\mathcal{H}_{2},\mathcal{H}_{3}$ where
$\mathcal{H}_{1}$ represents the processing part, $\mathcal{H}_{2}$ the memory
before recognition and $\mathcal{H}_{3}=\mathcal{H}_{2}$ the memory after
recognition. Throughout this paper we will have in mind this interpretation of
the Hilbert spaces $\mathcal{H}_{j}\left(  j=1,2,3\right)  .$ However, this is
just an illustration of what we are going to do, and the teleportation scheme
may be applied to very different situations.

We are mainly interested in the changes of the memory after the process of
recognition. For that reason we consider channels from the set of states on
$\mathcal{H}_{1}{\otimes}\mathcal{H}_{2}$ into $\mathcal{H}_{3}$. Main object
to be measured causing the recognition is here assumed to be a self-adjoint
operator
\[
F=\sum_{k,l=1}^{n}z_{k,l}F_{k,l}
\]
on $\mathcal{H}_{1}{\otimes}\mathcal{H}_{2}$ where the operators $F_{k,l}$ are
orthogonal projections (alternatively, we may take $F_{k,l}$ as an operator
valued measure). The channel $\Lambda_{k,l}$ describes the state of the memory
after the process of recognition if the outcome of the measurement according
to $F$ was $z_{k,l}$ and is given by
\[
\Lambda_{k,l}({\rho}{\otimes}{\gamma}):=\frac{\mathrm{Tr}_{1,2}(F_{k,l}%
{\otimes}{\mathbf{1}})({\rho}{\otimes}J{\gamma}J^{\ast}))(F_{k,l}{\otimes
}{\mathbf{1}})}{\mathrm{Tr}_{1,2,3}(F_{k,l}{\otimes}{\mathbf{1}})({\rho
}{\otimes}J{\gamma}J^{\ast}))(F_{k,l}{\otimes}{\mathbf{1}})}
\]
where ${\rho}$ and ${\gamma}$ (denoted $\rho_{O}$ above) are the state of the
processing part and of the memory before recognition and $J$ an isometry
extending from $\mathcal{H}_{2}$ to $\mathcal{H}_{2}{\otimes}\mathcal{H}_{3} $
and $\mathbf{1}$ denotes the identical operator. The value $\mathrm{Tr}%
_{1,2,3}(F_{k,l}{\otimes}{\mathbf{1}})({\rho}{\otimes\emph{J}{\gamma}%
\emph{J}^{\ast}})(F_{k,l}{\otimes}{\mathbf{1}})$ represents the probability to
measure the value $z_{k,l}$. So, obviously, we have to assume that this
probability is greater than 0. The state $\Lambda_{k,l}({\rho}{\otimes}%
{\gamma})$ gives the state of the memory after the process of recognition. The
elements of a basis $(b_{k})_{k=1}^{n}$ of $\mathcal{H}_{j}$ are interpreted
as elementary signals.

In this first attempt to our model described above, there appear still a lot
of effects being non-realistic for the process of recognition. Some examples
(cf. the last subsection) show that with this model one can describe extreme
cases such as storing the full information or total loss of memory, but - as
mentioned above - that is still far from being a realistic description.

In the paper \cite{FFO}, we restrict ourselves to finite dimensional Hilbert
spaces. Moreover, we assume equal dimension of the Hilbert spaces
$\mathcal{H}_{j}\left(  j=1,2,3\right)  $. It seems that infinite dimensional
schemes will lead to more realistic models. However, this is just a first
attempt to describe the brain function. Moreover, for finite dimensional
Hilbert spaces the mathematical model becomes more transparent and one can
obtain easily a general idea of the model. To indicate obvious generalizations
to more general situations and especially to infinite dimensional Hilbert
spaces we sometimes use notions and notations from the general functional
analysis \cite{FF1,FFL}.

\subsection{Basic Notions}

Let $\mathcal{H}_{1},\mathcal{H}_{2},\mathcal{H}_{3}$ be Hilbert spaces with
equal finite dimension:
\[
\dim\mathcal{H}_{j}=n,\;\;(j\in\{1,2,3\}).
\]
First we will represent these Hilbert spaces in a way that it seems to be
convenient for our considerations. Each of the spaces $\mathcal{H}%
_{1},\mathcal{H}_{2},\mathcal{H}_{3}$ can be identified with the space
${\mathbb{C}}^{n}$ of $n$-dimensional complex vectors. The space ${\mathbb{C}%
}^{n}$ again may be identified with the space $\{f:G\longrightarrow
\mathcal{C}\}$ of all complex-valued function on $G:=\{1,\ldots,n\}$. The
scalar product then is given by
\[
\langle f,g\rangle:=\sum_{k=1}^{n}\overline{\emph{f}\left(  k\right)
}g(k)=\int\overline{\emph{f}\left(  k\right)  }g(k)\mu(dk)
\]
where $\mu$ is the counting measure on $G$, i.e. $\mu=\sum_{k=1}^{n}\delta
_{k}$ with $\delta_{k}$ denoting the \textsc{Dirac} measure in $k$. So, each
of the spaces $\mathcal{H}_{j}$ can be written formally as an $L_{2}$-space:
\[
\mathcal{H}_{j}=L_{2}(G,\mu):=L_{2}(G)\;\;\;\;(j\in\{1,2,3\}).
\]
For the tensor product one obtains
\[
f{\otimes}g(k,l)=f(k)g(l)\;\;\;\;(f,g\in L_{2}(G),k,\in G),
\]
and we have
\[
\mathcal{H}_{1}{\otimes}\mathcal{H}_{2}=L_{2}(G\times G,\mu\times
\mu)=\mathcal{H}_{2}{\otimes}\mathcal{H}_{3}.
\]
We will abbreviate this tensor product by $L_{2}(G^{2},\mu^{2})$ or just by
$L_{2}(G^{2})$.

By $\mathcal{B}(\mathcal{H})$ we denote the space of all bounded linear
operators on a Hilbert space $\mathcal{H}$. In $\mathcal{B}(L_{2}(G))$ the
operator of multiplication by a function $g\in L_{2}(G)$ is given by
\[
(\mathcal{O}\sb{g}\,f)(k)=g(k)f(k)\;\;\;\;(f\in L_{2}(G),k\in G).
\]
Observe that for all $f,g\in L_{2}(G)$ one has
\[
\mathcal{O}_{f}\,g=\mathcal{O}_{g}\,f,\;\;\;\;\mathcal{O}_{f}^{\ast
}=\mathcal{O}_{\overline{f}}
\]
and for $f\in L_{2}(G)$ with $f(k)\not =0$ for all $k\in G$ it holds
$\mathcal{O}_{f}^{-1}=\mathcal{O}_{1/f}.$

The function $\mathbf{1}$, $\mathbf{1}(k)=1$ for all $k\in G,$ obviously
belongs to $L_{2}(G)$ and $\mathbf{1}=\mathcal{O}_{\mathbf{1}}$ is the
identity in $\mathcal{B}(L_{2}(G))$.

Consequently, an operator of multiplication $\mathcal{O}_{f}$ is unitary if
and only if $|f(k)|=1$ for all $k\in G$. \label{unitaer}

Further, we will use the mapping $J$ from $L_{2}(G)$ into $L_{2}(G^{2})$ given
by
\begin{equation}
(J\,f)(k,l)=f(k)\delta_{k,l}\;\;\;\;(f\in L_{2}(G),k,l\in G)\label{isometry}%
\end{equation}
where $\delta_{k,l}$ denotes the \textsc{Kronecker} symbol. It is immediate to
see that $J$ is an isometry. For the adjoint $J^{\ast}:L_{2}(G^{2}%
)\longrightarrow L_{2}(G)$ we obtain
\begin{equation}
(J^{\ast}\Phi)(k)=\Phi(k,k)\;\;\;\;(\Phi\in L_{2}(G^{2}),\;k\in
G).\label{isometry*}%
\end{equation}

Observe that $G$ equiped with the operation $\oplus:G\times G\longrightarrow
G,$ $k\oplus l:=(k+l)\mathrm{mod}~n$ \label{doplus} is a group. The operation
inverse to $\oplus$ we denote by $\ominus$. Let us remark that $k\ominus
l=k-l$ in the case $k>l$ and $k\ominus l=k-l+n$ if $k\leq l$. We conclude that
for all $k\in G$ the operator $U_{k}\in\mathcal{B}(L_{2}(G))$ given by
\begin{equation}
(U_{k}\,f)(m):=f(k\oplus m)\;\;\;\;(f\in L_{2}(G))\label{U_k}%
\end{equation}
is unitary.

Now, let $(b_{k})_{k=1}^{n}$ be an orthonormal basis in $L_{2}(G)$, and denote
by $(B_{k})_{k=1}^{n}$ the sequence of multiplication operators corresponding
to the elements of this basis, i.e. $B_{k}:=\mathcal{O}_{b_{k}},k\in G.$ Then
for $k,l\in G$ we put
\begin{equation}
\label{xikl}\xi_{k,l}:=(B_{k}{\otimes} U_{l})J\,\mathbf{1}.
\end{equation}
One can show that the sequence $(\xi_{k,l})_{k,l\in G}$ is an orthonormal
basis in $L_{2}(G^{2}).$ And we denote by $F_{i,j}\in\mathcal{B}(L_{2}%
(G^{2}))$ the projection onto $\xi_{i,j}$, i.e.
\begin{equation}
F_{i,j}:=|\xi_{i,j}\rangle\langle\xi_{i,j}|=\langle\xi_{i,j},\cdot\rangle
\xi_{i,j}.\label{fij}%
\end{equation}

\subsection{Channels}

\begin{definition}
\label{dentstate} Let ${\gamma}$ be a state on $\mathcal{H}_{2}=L_{2}(G)$
(i.e. ${\gamma}$ is a positive trace-class operator with $\mathrm{Tr}({\gamma
})=1$). The state $\mathbf{e}({\gamma})$ on $L_{2}(G^{2})=\mathcal{H}%
_{2}{\otimes} \mathcal{H}_{3}$ given by
\begin{equation}
\label{entstate}\mathbf{e}({\gamma})=J{\gamma} J^{*}%
\end{equation}
where $J$ is the isometry given by (\ref{isometry}) we call the entangled
state corresponding to ${\gamma}$.
\end{definition}

Now, let ${\rho}$ and ${\gamma}$ be states on $\mathcal{H}_{1}$ resp.
$\mathcal{H}_{2}$, the state $\mathbf{e}({\gamma})$ (usually denoted by
$\sigma$ \cite{FO1}) will be a state on $\mathcal{H}_{2}{\otimes}%
\mathcal{H}_{3}.$ . Remember that we assumed $\mathcal{H}_{1}=\mathcal{H}%
_{2}=\mathcal{H}_{3}=L_{2}(G)$. The numbering only indicates the meaning of
the states (we recall that $\mathcal{H}_{1}$ represents the processing part,
$\mathcal{H}_{2}$ the memory before and $\mathcal{H}_{3}$ the memory after the
recognition process.) Then ${\rho}{\otimes}\mathbf{e}({\gamma})$ is a state on
$\mathcal{H}_{1}{\otimes}\mathcal{H}_{2}{\otimes}\mathcal{H}_{3}$ and we
observe immediately
\begin{equation}
{\rho}{\otimes}\mathbf{e}({\gamma})=({\mathbf{1}}{\otimes}J)({\rho}{\otimes
}{\gamma})({\mathbf{1}}{\otimes}J^{\ast}).\label{zustandgross}%
\end{equation}
In subsection \ref{generalization} we calculate explicitly the trace of
\begin{equation}
(F_{i,j}{\otimes}{\mathbf{1}})({\rho}{\otimes}\mathbf{e}({\gamma}%
))(F_{i,j}{\otimes}{\mathbf{1}})=(F_{i,j}{\otimes}{\mathbf{1}})({\mathbf{1}%
}{\otimes}J)({\rho}{\otimes}{\gamma})({\mathbf{1}}{\otimes}J^{\ast}%
)(F_{i,j}{\otimes}{\mathbf{1}}).\label{formel fij1}%
\end{equation}
The following proposition will be very useful for this.

\begin{proposition}
\cite{FFO} \label{satz1} Let $(g_{k})_{k=1}^{n}$ and $(h_{k})_{k=1}^{n}$ be
orthonormal systems in $L_{2}(G)$ and ${\rho}$ and ${\gamma}$ states on
$L_{2}(G)$ having the following representations:
\begin{align*}
{\rho}  & =\sum_{k=1}^{n}{\alpha}_{k}|g_{k}><g_{k}|,\hspace{1cm}{\gamma}%
=\sum_{k=1}^{n}{\beta}_{k}|h_{k}><h_{k}|,\\
{\alpha}_{k}  & \geq0,{\beta}_{k}\geq0,\sum_{k=1}^{n}{\alpha}_{k}=\sum
_{k=1}^{n}{\beta}_{k}=1.
\end{align*}
Then for all $i,j\in G$
\begin{equation}
(F_{i,j}{\otimes}{\mathbf{1}})({\rho}{\otimes}\mathbf{e}({\gamma}%
))(F_{i,j}{\otimes}{\mathbf{1}})=F_{i,j}{\otimes}\sum_{k,l=1}^{n}{{\alpha}%
_{k}}{\beta}_{l}|G_{i,j}g_{k}{\otimes}h_{l}><G_{i,j}g_{k}{\otimes}%
h_{l}|.\label{formelsatz1}%
\end{equation}
where $G_{i,j}$ is given by, for $i,j\in G$
\begin{equation}
\hspace{-3cm}G_{i,j}:=J^{\ast}(U_{j}{\otimes}{\mathbf{1}})({B}_{i}^{\ast
}{\otimes}{\mathbf{1}})=J^{\ast}(U_{j}B_{i}^{\ast}{\otimes}{\mathbf{1}%
})\label{gij}%
\end{equation}
where ${B}_{i}^{\ast}=\mathcal{O}_{b_{i}}^{\ast}=\mathcal{O}_{\overline{{b}%
}_{i}}.$
\end{proposition}

\label{channels} Denote by $\mathcal{T}$ the set of all positive trace-class
operators on $L_{2}(G)$ including the null operator $\mathbf{0}$,
\[
\mathbf{0}(f)=0\hspace{3cm}(f\in L_{2}(G)).
\]
We fix an operator $\tau\in\mathcal{T}$ having the representation
\begin{equation}
\label{darst1}\tau=\sum_{k=1}^{n}\gamma_{k}| h_{k}>< h_{k}|
\end{equation}
with $(\gamma_{k})_{k\in G}\subseteq[0,\infty)$ and $(h_{k})_{k\in G}$ being
an orthonormal basis in $L_{2}(G)$.

The linear mapping $K_{\tau}: \mathcal{T}\longrightarrow\mathcal{T}$ given by
\begin{equation}
\label{ktau}K_{\tau}({\rho}):=\sum_{k=1}^{n}\gamma_{k}\mathcal{O}_{ h_{k}%
}{\rho} \mathcal{O}_{h_{k}}^{*}\hspace{3cm}({\rho}\in\mathcal{T})
\end{equation}
depends only on the operator $\tau$ but not on its special representation.

\begin{definition}
\label{dchannel} Denote by $\mathcal{S}$ the set of all states on $L_{2}(G)$
and for $\tau\in\mathcal{T}$ by $\mathcal{S}_{\tau}$ the set of all states
${\rho}$ from $\mathcal{S}$ with the property that $\mathrm{Tr}K_{\tau}({\rho
})$ is positive:
\begin{equation}
\label{stau}\mathcal{S}_{\tau}:=\{{\rho}\in\mathcal{S}:\mathrm{Tr}K_{\tau
}({\rho})>0\}.
\end{equation}
For $\tau\in\mathcal{T}$ the mapping $\hat{K_{\tau}}:\mathcal{S}_{\tau
}\longrightarrow\mathcal{S}$ given by
\begin{equation}
\label{channel}\hat{K_{\tau}}({\rho}):=\frac{1}{\mathrm{Tr}K_{\tau}({\rho}%
)}K_{\tau}({\rho})\hspace{2cm}({\rho}\in\mathcal{S}_{\tau})
\end{equation}
is called the \textsc{channel} corresponding to $\tau$. The channel
corresponding to $\tau$ is called \textsc{unitary} if there exists an unitary
operator $U$ on $L_{2}(G)$ such that $\hat{K_{\tau}}({\rho})=U{\rho} U^{*}$
\end{definition}

Observe that the channel $\hat{K_{\tau}}$ is in general nonlinear.

Let us make some remarks on the physical meaning of the channels ${K_{\tau}}$
and $\hat{K_{\tau}}$. The channels ${K_{\tau}}$ are mixtures of linear
channels of the type
\begin{align}
K^{h}(\rho):=\mathcal{O}_{h}\rho\mathcal{O}_{h}^{*}\hspace{2cm}(\rho\in)
\end{align}
with $h\in L_{2}(G),\ \Vert h \Vert=1.$ Let us consider the more general case
\begin{align}
\Vert h\Vert>0,\ |h(k)|\le1\hspace{2cm}(k\in G).
\end{align}
We define an operator $t_{h}:L_{2}(G)\longrightarrow L_{2}(\{1,2\}\times G)$
by setting for all $f\in L_{2}(G)$ and $k\in G$
\begin{align*}
(t_{h}\, f)(l,k)=\left\{
\begin{array}
[c]{ll}%
h(k)f(k) & \mbox{  for  }l=1\\
& \\
\sqrt{1-|h(k)|^{2}}f(k) & \mbox{  for }l=2.
\end{array}
\right.
\end{align*}
The operator $t_{h}$ is an isometry from $L_{2}(G)$ to $L_{2}(\{1,2\}\times
G)\cong L_{2}(\{1,2\})\otimes L_{2}(G).$ Indeed,
\begin{align}
||t_{h}\, f||^{2}=\sum_{l=1}^{2}\sum_{k=1}^{n}|t_{h}\,f(l,k)|^{2}=\sum
_{k=1}^{n}\big(|h(k)|^{2}+1-|h(k)|^{2}\big)|f(k)|^{2}=||f||^{2}.
\end{align}
Consequently, the mapping $E_{h}: \mathcal{B}(L_{2}(\{1,2\}\times G))
\longrightarrow\mathcal{B}(L_{2}(G))$ given by
\[
E_{h}(B):=t_{h}^{*}Bt_{h}
\]
is completely positive and identity preserving. The channel $E_{h}^{*}%
(\rho)=t_{h}\rho t_{h}^{*}~$ is the corresponding linear channel from the set
of states on $L_{2}(G)$ into the set of states on $L_{2}(\{1,2\}\times G).$
The space $L_{2}(\{1,2\}\times G)$ has an orthogonal decomposition into
$L_{2}(\{1\}\times G)$ and $L_{2}(\{2\}\times G)$ both being trivially
isomorphic to $L_{2}(G)$. Performing a measurement according to the projection
onto $L_{2}(\{1\}\times G)\cong L_{2}(G)$ given the state $E_{h}^{*}(\rho)$
one obtains the state $\hat{K}^{h}(\rho).$ A measurement according to the
projection onto $L_{2}(\{2\}\times G)\cong L_{2}(G)$ leads to the state
$\hat{K}^{\sqrt{1-|h|^{2}}}(\rho)$.

\subsection{The State of the Memory after Recognition}

\label{generalization} Let us recall that for states ${\rho},{\gamma}$ on
$L_{2}(G)$ and $i,j\in G$
\[
(F_{i,j}{\otimes}{\mathbf{1}})({\rho}{\otimes}\mathbf{e}({\gamma}%
))(F_{i,j}{\otimes}{\mathbf{1}})
\]
is a linear operator from $L_{2}(G^{3})$ into $L_{2}(G^{2}),$ and that (cf.
(\ref{formel fij1})) it is equal
\[
(F_{i,j}{\otimes}{\mathbf{1}})({\mathbf{1}}{\otimes}J)({\rho}{\otimes}{\gamma
})({\mathbf{1}}{\otimes}J^{\ast})(F_{i,j}{\otimes}{\mathbf{1}}).
\]
In the following we consider the family of channels $(\Lambda_{i,j})_{i,j\in
G}$ from the set of product states ${\rho}{\otimes}{\gamma}$ on $\mathcal{H}%
_{1}{\otimes}\mathcal{H}_{2}$ into the states on $\mathcal{H}_{3}$ given by
\begin{equation}
\Lambda_{i,j}({\rho}{\otimes}{\gamma}):=\frac{\mathrm{Tr}_{1,2}(F_{i,j}%
{\otimes}{\mathbf{1}})({\rho}{\otimes}\mathbf{e}({\gamma}))(F_{i,j}{\otimes
}{\mathbf{1}})}{\mathrm{Tr}_{1,2,3}(F_{i,j}{\otimes}{\mathbf{1}})({\rho
}{\otimes}\mathbf{e}({\gamma}))(F_{i,j}{\otimes}{\mathbf{1}})}%
\label{dlambdaij}%
\end{equation}
where $\mathrm{Tr}_{1,2}$ resp. $\mathrm{Tr}_{1,2,3}$ denotes the partial
trace with respect to the first two components resp. the full trace with
respect to all three spaces. In the sequel we always will assume that
\begin{equation}
\mathrm{Tr}_{1,2,3}(F_{i,j}{\otimes}{\mathbf{1}})({\rho}{\otimes}%
\mathbf{e}({\gamma}))(F_{i,j}{\otimes}{\mathbf{1}})>0.\label{>0}%
\end{equation}
Let ${\rho}$ and ${\gamma}$ are given as in Proposition \ref{satz1}. Since
$(\xi_{i,j})_{i,j\in G}$ is an orthonormal basis in $L_{2}(G^{2})$ we get from
Proposition \ref{satz1}
\begin{equation}
\mathrm{Tr}_{1,2}(F_{i,j}{\otimes}{\mathbf{1}})({\rho}{\otimes}\mathbf{e}%
({\gamma}))(F_{i,j}{\otimes}{\mathbf{1}})=\sum_{k,l=1}^{n}{{\alpha}_{k}}%
{\beta}_{l}\langle G_{i,j}g_{k}{\otimes}h_{l},\cdot\rangle\ G_{i,j}%
g_{k}{\otimes}h_{l}%
\end{equation}

Summarizing, we get the following representation of $\Lambda_{i,j}$:

\begin{proposition}
\cite{FFO} \label{satz5.1} Let ${\rho}$ and ${\gamma}$ be given as in
Proposition \ref{satz1}. Further, assume (\ref{>0}). Then
\begin{equation}
\label{slambdaij}\Lambda_{i,j}({\rho}{\otimes}{\gamma})=\frac{\sum_{k,l=1}%
^{n}{\alpha}_{k}{\beta}_{l} \langle G_{i,j}g_{k}{\otimes} h_{l},\cdot
\rangle\ G_{i,j}g_{k}{\otimes} h_{l}}{\sum_{k,l=1} ^{n}{\alpha}_{k} {\beta
}_{l} || G_{i,j}g_{k}{\otimes} h_{l}||^{2}}%
\end{equation}
where for $\Phi\in\mathcal{L}_{2}(G^{2})$
\begin{equation}
|| G_{i,j}\Phi||^{2}=\sum_{m=1}^{n}|b_{i}|^{2}(m\oplus j)|\Phi(m\oplus
j,m)|^{2}.
\end{equation}

\end{proposition}

Fortunately, we can find expressions for the state $\Lambda_{i,j}({\rho
}{\otimes}{\gamma})$ of the memory after the recognition process being in many
cases simpler. We can express the teleportation channel $\Lambda_{i,j}$ with
the help of the channels $K_{\tau}$ we introduced in the previous subsection.

\begin{proposition}
\cite{FFO}\label{theorem} Let $i,j\in G$ and let ${\rho}$ be a state from
$\mathcal{S}_{|\overline{{b_{i}}}><\overline{{b_{i}}}|}$ (cf. (\ref{ktau}) and
Definition \ref{dchannel}). Further, let ${\gamma}$ be a state from
$\mathcal{S}$ such that
\begin{equation}
U_{j}K_{|\overline{{b_{i}}}><\overline{{b_{i}}}|}({\rho})U_{j}^{\ast}%
\in\mathcal{S}_{{\gamma}}.\label{ujk}%
\end{equation}
Then
\begin{equation}
\Lambda_{i,j}({\rho}{\otimes}{\gamma})=\hat{K}_{{\gamma}}\circ K^{j}\circ
\hat{K}_{|\overline{{b_{i}}}><\overline{{b_{i}}}|}({\rho})\label{lambdaij2}%
\end{equation}
where $K^{j}$ denotes the unitary channel given by $K^{j}({\rho})=U_{j}{\rho
}U_{j}^{\ast}.$
\end{proposition}

\begin{remark}
All proofs of this paper can be seen in \cite{FFO}.
\end{remark}

\textbf{Concluding remarks:} We touched the problem of finding simplified
models for the recognition process. We were interested in how the input signal
arriving at the brain is entangled (connected) to the memory already stored
and the consciousness that existed in the brain, and how a part of the signal
will be finally stored as a memory. It is clear that this simple model is just
for illustration and can not serve for describing realistic aspects of
recognition. Choosing a more complex basis one obtains expressions depending
heavily on the states ${\rho}$ and ${\gamma}$. Though the above presented
model is only a first attempt it shows that there are possibilities to model
the process of recognition. To get closer to realistic models we will try to
refine the above models by

\begin{itemize}
\item[-] passing over to infinite Hilbert spaces,

\item[-] replacing pure states by coherent states on the Fock space,

\item[-] making more complex measurements than simple one-dimensional
projections $F_{i,j}$,

\item[-] replacing the trivial entanglement $J$ by a more complex one based on
beam splitting procedures, and finally

\item[-] examing whether some symmetry breaking as in \cite{ES} will occur in
the process of recognition and storing memory.
\end{itemize}

\end{document}